\documentclass[prb,aps,twocolumn,epsfig]{revtex4}
\usepackage{graphicx}
\usepackage{dcolumn}
\usepackage{bm}

\begin{document}
\title {A simple kinetic sensor to structural transitions}

\author{U. Chandni\footnote[1]{email:chandni@physics.iisc.ernet.in}}
\author{Arindam Ghosh}
\address{Department of Physics, Indian Institute of Science, Bangalore 560 012, India}

\begin{abstract}
Driven non-equilibrium structural phase transformation has been probed using time varying resistance fluctuations or noise. We demonstrate that
the non-Gaussian component (NGC) of noise obtained by evaluating the higher order statistics of fluctuations, serves as a simple kinetic
detector of these phase transitions. Using the martensite transformation in free-standing wires of nickel-titanium binary alloys as a prototype,
we observe clear deviations from the Gaussian background in the transformation zone, indicative of the long range correlations in the system as
the phase transforms. The viability of non-Gaussian statistics as a robust probe to structural phase transition was also confirmed by comparing
the results from differential scanning calorimetry measurements. We further studied the response of the NGC to the modifications in the
microstructure on repeated thermal cycling, as well as the variations in the temperature drive rate, and explained the results using established
simplistic models based on the different competing time scales. Our experiments (i) suggest an alternative method to estimate the transformation
temperature scales with high accuracy, and (ii) establish a connection between the material-specific evolution of microstructure to the
statistics of its linear response. Since the method depends on an in-built long-range correlation during transformation, it could be portable to
other structural transitions, as well as to materials of different physical origin and size.
\end{abstract}


\maketitle

\section{Introduction}
Structural transitions constitute an evolving field of research comprising of unique and interesting mechanisms like creep and rupture
\cite{miguel}, plastic deformation \cite{dimiduk,csikor}, externally induced crystal structure transformations \cite{bradby}, the shape memory
effect and pseudoelasticity \cite{otsuka} and so on. The revolution in material research, especially in nanoscale functional materials,
ferromagnetic shape memory alloys etc., have added, both in complexity and diversity, to these transitions, particularly in the nature of their
response to external stimuli such as mechanical stress, temperature, current or magnetic field.

A commonness in structural transitions in diverse materials lies in the fact that the quenched disorder at the micro scale often manifests in
discontinuities, avalanches or bursts and step changes in many of the physical responses. A description of the response within the framework of
continuum models, used for instance, to describe the viscous flow at the onset of plasticity, breaks down, thereby demanding newer approaches to
study these systems \cite{devincre,mo}. The crystallographic constraints and long range interactions lead to collective motions and
avalanche-like behavior that have been the major focus of many recent works, indicating a universality in the statistical mechanics of these
processes \cite{carrillo,kuntz,durin}. They form a subset of a larger class of systems exhibiting non-equilibrium dynamics
\cite{sethna1,sethna2}. These systems are found to posses a complex free energy landscape, where metastable states are separated by large
potential energy barriers.

Avalanches play a major role in driven non equilibrium systems manifesting in various forms like Barkhausen noise in magnetic
hysteresis\cite{durin}, acoustic emission under mechanical stress\cite{weiss} and so on. It has been observed that for such ``crackling" noise,
avalanches lead to fluctuations of physical properties that often become scale invariant or self similar. However, the statistics of these
fluctuating dynamic quantities have largely been confined to the study of universality in the underlying physical processes, for example, in
self organization of disorder\cite{reche2}, but hardly exploited as a characterizing tool for material dependent transition properties
themselves. Here, we aim to use higher order statistics of fluctuations as a sensor to structural transitions and the microstructural
modifications in a typical driven-nonequilibrium system, namely, the martensite transformation.

Martensite transformation has been known to be a prototype of out-of-equilibrium phenomena\cite{reche} and at the same time constitutes the
basis of a large variety of practical applications including actuators, micro-pumps and so on in material systems, that display shape memory
effect and pseudoelasticity \cite{otsuka}. These transitions are known to be athermal, as large elastic energy barriers that separate the
metastable states require a driving force in the form of time varying stress or temperature for the transition to proceed, and cannot be driven
by thermal fluctuations alone \cite{reche}. Interplay of disorder and long range elastic forces is known to play a significant role in the
transformation kinetics and it is this feature that we aim to exploit in the present study.

\begin{figure}
\begin{center}
\includegraphics[width=10cm,height=8cm]{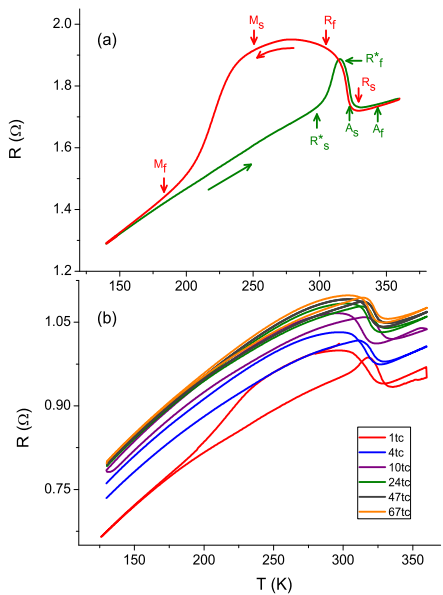}
\end{center}
\vspace{-0.7cm} \caption{(Color Online) (a) A typical resistance vs temperature plot exhibited by equiatomic Ni-Ti alloys. The transition
temperature scales are marked according to common practices. (b) Resistance vs temperature curves for the NiTi wire for various thermal cycles
(tc).} \label{figure1}
\end{figure}

The kinetics of martensitic transformation has been previously studied using acoustic emission measurements which probe the size distribution
and duration of avalanches and their universal features \cite{carrillo,planes}. However, the transformation temperature scales and the
microstructural evolution which are also essential for many device applications are addressed only in limited number of cases~\cite{reche}.
Generally, one resorts to the conventional methods of calorimetry or standard resistivity measurements for the determination of these parameters
\cite{otsuka}. Apart from inherent inaccuracies and subjectivities involved, many of these techniques fail as one scales down the size and
amount of materials \cite{juan}, becoming mostly inapplicable in the nanoscale systems. On the other hand, electrical noise, or fluctuations in
resistivity, has also shown distinct and interesting results in martensite transformation and avalanche dynamics~\cite{acta,prl}, and could be a
potential tool to characterize the transition. Indeed, it was recently shown that higher order avalanche statistics is an excellent probe to
martensite transformation, where detection of structural change depends on in-built long range correlation, rather than the amount of energy
released \cite{prl}. However, sensitivity of such a technique to small scale microstructural change is still not known.

The present study is hence relevant in three ways. First, it provides a kinetic detector which could detect both magnitude and changes in the
transformation temperatures as the material undergoes any slight modification in its structural morphology. Although resistivity measurements
have been used for over three decades \cite{otsuka} as a tool next to the calorimetric measurements, we show here that the phase transformation
start and finish temperatures assigned based on the resistivity curves often lead to difficulties. We aim to establish that the fluctuation in
resistance and its higher order statistics could reveal more detailed and accurate information about the transformation parameters as compared
to conventional resistivity measurements. Second, though the athermal dynamics and the avalanche statistics have been well established in the
context of crackling behavior\cite{sethna2,carrillo,vives}, a clear link between this statistical approach and the microstructural details of
these materials is lacking. Statistics of fluctuations \cite{prl} could again be used to probe this missing link, by altering the microstructure
of the sample through thermal cycling as well as by varying the drive rate. This is crucial to many applications of shape memory alloys since a
minute change in the temperature scales on repeated use could result in major consequences in their applications. Third, even though the
martensitic transformation is expected to be associated with a long-range elastic interaction, a direct probe to the nature and effect of this
interaction is still absent. Here, we address this issue with higher-order fluctuation statistics as the experimental tool. A connection between
the statistical parameters and the material specific microstructural details that our results indicate might also be applicable to many forms of
structural transition from a common platform. In addition, although the various timescales have been applied to explain the overlap of
avalanches during the phase transformation, a clear link between these timescales and the material properties is not well established. We
propose a simplistic model which accounts for the variation of the different transition parameters with increase in drive rate.

 Sections II and III of the paper summarize the material and the experimental methods respectively. In section IV, we present the major results and in section V, we
present a brief discussion and conclude.

\section{Material}
The sample used was a nickel-titanium (NiTi) wire of rectangular cross-section
$0.0005''\times 0.002''$ procured from M/s Fortwayne Metals Inc. which was
straight annealed and quenched in a strand furnace. The wire was found to show
a two stage phase transformation from austenite (cubic B2) to R-phase
(rhombohedral) and R-phase to martensite (B19/B19' monoclinic) phase on cooling
and vice versa on heating \cite{otsuka}. The wire was swept through a
temperature range of 360~K-120~K for over 50 times, so that defect levels
reached a stable configuration. Resistivity fluctuations or noise measurements
were carried out at appropriate thermal cycles so that a correspondence could
be drawn with the changes in microstructure. The drive rate dependence was
studied only after 75 thermal cycles so that the effects of cycling and driving
induced changes in the sample morphology could be decoupled.

\section{Experimental Methods}
\subsection{Time averaged resistivity}
Resistivity measurements were carried out on the wire samples using a four
probe geometry. A typical resistivity curve is as shown in Fig.1 (a). The
resistivity exhibits a hysteresis on cooling and heating indicative of the
shape memory effect. The transformation temperatures can be extracted using
conventional formulation \cite{otsuka} as is shown. The wire exhibits a two
stage phase transformation with an intermediate R-phase. The subscripts $s$ and
$f$ denote the start and finish temperature scales of the different phases.
\subsection{Time dependent resistivity}
The sample was current biased using a low noise source and the voltage fluctuations were measured over an interval of time. We have used a five
probe geometry and a dynamically balanced wheatstone's bridge arrangement (Fig.2) to measure the noise as proposed by Scofield \cite{scofield}.

\emph{\textbf{Power spectral density (PSD)}}

The power spectrum $S(f)$ quantifies the noise in the frequency domain. If $\rho(t)$ is the fluctuating resistivity, the power spectrum is
defined as,

\begin{equation}
S_\rho(f)=\lim_{T\rightarrow \infty}\left(\frac{1}{2T}\right) \left(\int_{-T}^{T} dt \Delta \rho(t) e^{-i2\pi{}ft}\right)^{2}
\end{equation}

Large classes of fluctuations in condensed matter systems show a power spectrum that has 1/f$^{\alpha}$ dependence, with $\alpha\approx{1}$ and
hence called $1/f$ noise.

In a typical experiment, the voltage fluctuations are acquired at a sampling rate of 1000~Hz. The data is subsequently decimated using a three
stage decimation algorithm with appropriate anti-aliasing at each stage, which removes the unwanted signals. Care has been taken to remove the
extraneous noise in the system, using appropriate grounding and shielding using a faraday cage. An in depth review of the experimental setup and
the measurement scheme including the basic digital signal processing steps could be found in our earlier paper \cite{acta}.

\begin{figure}
\begin{center}
\includegraphics[width=5cm,height=6cm]{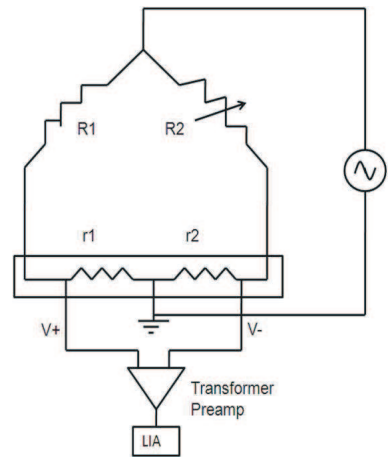}
\end{center}
\vspace{-0.7cm} \caption{(Color Online) Schematic of the dynamically balanced wheatstone's bridge arrangement. R1 and R2 constitute the current
limiting and balancing resistors respectively and r1 and r2 constitute the sample with the middle contact grounded. The voltage is measured
across a transformer preamplifier and then connected to a lock-in amplifier.}
\end{figure}

\emph{\textbf{Second spectrum and the non-Gaussian component}}

Along with the power spectral density, we have also evaluated the higher order statistics of fluctuations in the form of second spectrum. The
second spectrum is a fourth order moment of the fluctuations, that detects the weak correlations in the noise generating mechanism. We have used
the approach followed by Seidler and Solin \cite{seidler}, which is in turn a modification of the approach followed by Restle et al.
\cite{restle}. The resistivity fluctuations over a period of time, $\Delta\rho$(t), are recorded and then digitally bandpass filtered over a
frequency band ($f_L$, $f_H$). It is then squared point by point and the power spectrum of the time dependent variance of the noise signal is
computed which is called the second spectrum, expressed analytically as
\begin{equation}
S^{(2)}(f)=\int_{0}^{\infty} \langle
\Delta\rho^{2}(t)\Delta\rho^{2}(t+\tau)\rangle_{t} \cos{(2\pi f\tau)}d\tau
\end{equation}

The second spectrum was computed for each of these data sets, over a frequency band of $f_L=1$Hz and $f_H=3$Hz, and was normalized with the
square of the corresponding power spectrum, $S_\rho(f)$, as

\begin{equation}
s^{(2)}(f) = S^{(2)}(f)/[\int_{f_L}^{f_H}S_\rho(f)df]^2
\end{equation}
The Gaussian background has been calculated using the expression evaluated in Ref.22:

\begin{equation}
s^{(2)}(f)=\frac{2}{f}\frac{\ln(f_H-f)(f_L+f)/f_H f_L}{\ln[(f_H f_L)]^2}
\end{equation}
Even weak correlations in the stochastic process contributes a non-Gaussian component (NGC) to the fluctuations, which can be readily identified
from the deviation from its expected Gaussian background. In some cases, the NGC exhibits frequency dependence, as is clearly shown in Ref.18.
Finite detection bandwidth sets the limit of sensitivity of $s^{(2)}(f)$ to non-Gaussian fluctuations. For comparison we calculated the total
fourth order noise as,
\begin{equation}
\sigma^{(2)} = \int_0^{f_H-f_L}s^{(2)}(f)df
\end{equation}

Within a detection bandwidth of $f_L=1$~Hz and $f_H=3$~Hz, assuming $S_\rho(f)\approx 1/f$, the value of $\sigma^{(2)}$ for a Gaussian
background should be around 0.93 in accordance with equation(4). In our case, we see that when correlations are not significant, we obtain a
value close to 0.93 as the noise is approximately 1/f in nature over most of the temperature region. Any deviation from this Gaussian background
can be considered to be the NGC, which is the signature of long-range correlations in the system.

\section{Results}
\subsection{Resistivity fluctuations and statistics}

We first probed the signature of avalanches in the resistivity fluctuations across the transformation zone (using temperature or stress in the
case of martensite transformation). Since the martensite transformation in NiTi is primarily athermal, the avalanches are induced only by a
continuous sweeping of $T$. The characteristics of transformation, and hence the statistics of fluctuations, depend crucially on the sweep rate
$r = dT/dt$, and will be discussed in detail later. Fig. 3(a) shows the time-dependence of resistivity around $T = 250$~K for a NiTi-device
cooled at $r = 1$~K/min. It is clear that the fluctuations are much smaller in the static condition, {\it i.e.} keeping $T$ constant at the same
value, when avalanches are expected to be absent.

In a typical experiment, the fluctuations in $\rho$ were measured with $T$ being swept from $T \gg A_f \sim 325$~K to $T \ll M_f \sim 180$~K
($A_f$ and $M_f$ are the Austenite finish and Martensite finish temperature scales respectively), with the T-range divided in equal time windows
of 8.74 minutes. Both PSD and higher order statistics of noise were subsequently calculated in each window and plotted against the temperature
corresponding to its midpoint. The PSD of the fluctuations varied as $S_\rho(f) \sim 1/f^\alpha$ over the experimental bandwidth, where $\alpha
\sim 1.1-1.3$ (See Fig.~3b). The magnitude of $\alpha$ was found to be slightly smaller than that in NiTi thin films observed earlier, which
could be due to greater inhomogeneity in the bulk NiTi system in the present case. The PSD magnitude was found to be $T$-dependent as well,
increasing monotonically with $T$ in all devices over the observed $T$ range (See Figs.6(a) and 8(a)).

The behavior of the second spectrum was found to be different. Typically, $s^{(2)}$ agrees with the Gaussian background (the dashed line in
Fig.~3c), except at certain values of $T$ where it was found to be orders of magnitude higher, indicating significant non-Gaussian contribution
to noise. Fig.~3c illustrates this for a few representative $T$, where $s^{(2)}$ at various temperatures $T < 240K$ is not only much higher than
the Gaussian background, but also develops a weak slope $s^{(2)} \sim 1/f^\beta$ ($\beta \approx 0.4-0.6$). The absolute magnitude of $s^{(2)}$
was however found to be highly sample dependent, without any clear trend, but its variation with $T$ was qualitatively similar in all devices,
which we shall now focus on. No non-Gaussian component of noise could be detected in the static mode (not shown), implying non-Gaussian behavior
to arise from correlated avalanches (not shown).

Unlike PSD, we find the NGC to be non-monotonic in $T$, and manifest as strong peaks in $\sigma^{(2)}$ at specific temperatures. One expects
four structural transitions in NiTi within this range: Austenite to R-phase (R$\leftarrow$A) and R-phase to martensite (M$\leftarrow$R) during
cooling, and martensite to R*-phase (M$\rightarrow$R*) and R*-phase to austenite (R*$\rightarrow$A) during heating. In Fig.~4(a) (cooling) and
4(c) (heating) these four transitions are identified by the latent heat release in the differential scanning calorimetry measurements (Fig.~4(b)
and 4(d)). Strikingly, the peaks in $\sigma^{(2)}$ appear primarily around these phase transformation zones. Similar correlation of non-Gaussian
electrical noise and first order structural phase change was observed in NiTi thin film as well~\cite{prl}, outlining an excellent {\it kinetic}
detector of structural phase transition. The NGC in resistivity noise has been attributed to emergence of a long range correlation in many
systems with electronic~\cite{swastik}, structural~\cite{prl} or magnetic\cite{hardner} phase transitions. The magnitude of $\sigma^{(2)}$ is
directly associated to a correlation length $\xi$ which, in case of martensitic alloys, characterizes the range of elastic correlation in the
system. As the transition is approached, the avalanches become more frequent, which overlap in space, as well as in time in the case of faster
sweeps, leading to strongly hierarchical and cooperative atomic kinetics. The non-Gaussian fluctuations in $\rho$ is a result of electron wave
function coupling to the size of atomic avalanche, {\it i.e.} the number of atoms undergoing displacive movement. Consequently, $\xi$ may attain
a maximum at a critical temperature ($T_c$) where the avalanche overlap is also most probable. This would occur when maximum amount of parent
phase is transformed into the product phase, as confirmed by the agreement of $T_c$ with the peak heat release/absorb temperature scales in DSC
measurements.

In the following, we stress on two specific questions: (1) Can NGC provide a convenient scheme to evaluate the transformation temperatures, such
as the $M_f$, $A_f$ etc? Secondly (2) how do small scale microstructural modifications influence the non-Gaussian characteristics? The
microstructural evolution of the material depends mainly on two processes: Thermal cycling, which changes the local disorder level, and also the
drive rate, which leads to a significant modification in the internal stress levels and avalanche overlap. To analyze these cases in detail, we
shall focus mostly on a specific transformation zone, namely that from R-phase to the martensite phase (M$\leftarrow$R) while cooling, which in
most devices extend from $T \sim 180$~K$\leftarrow 240$~K.

\begin{figure}
\begin{center}
\includegraphics[width=9cm,height=8cm]{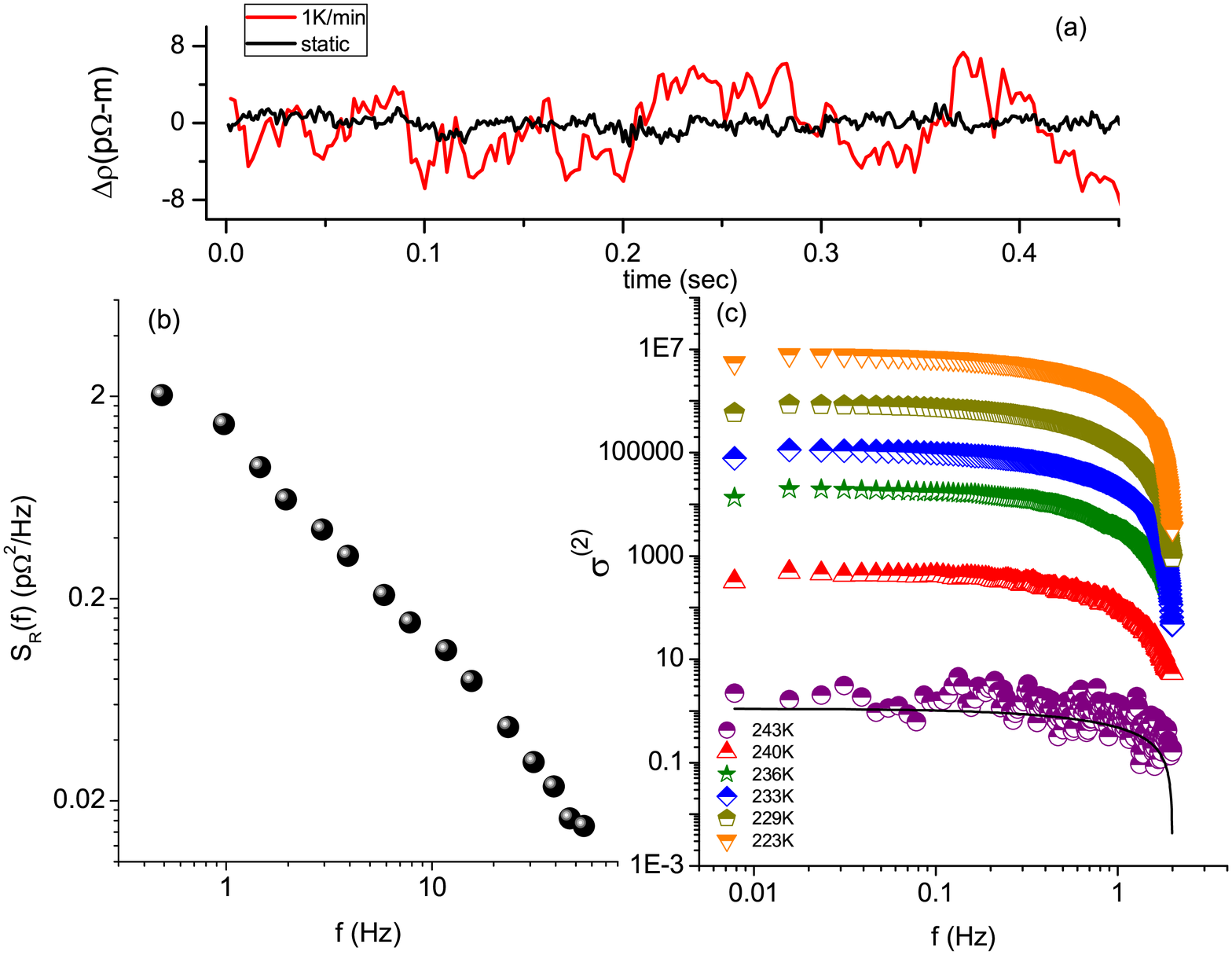}
\end{center}
\vspace{-0.7cm} \caption{(Color Online)(a) A typical time series of the resistance fluctuations while the NiTi sample was being ramped at 1K/min
around a mean temperature of 250~K and while keeping the temperature constant at 250~K (b) A typical power spectrum showing the $1/f^\alpha$
behavior. (c) The normalized second spectra for various temperatures during a cooling run. The increase in the integrated second spectra, which
ultimately leads to a peak in $\sigma^{(2)}$ is seen as a shift in the spectra, in this plot. The Gaussian background is plotted as a dotted
line.} \label{figure2}
\end{figure}

\subsection{Transition temperatures}

\begin{figure}
\begin{center}
\includegraphics[width=8cm,height=7cm]{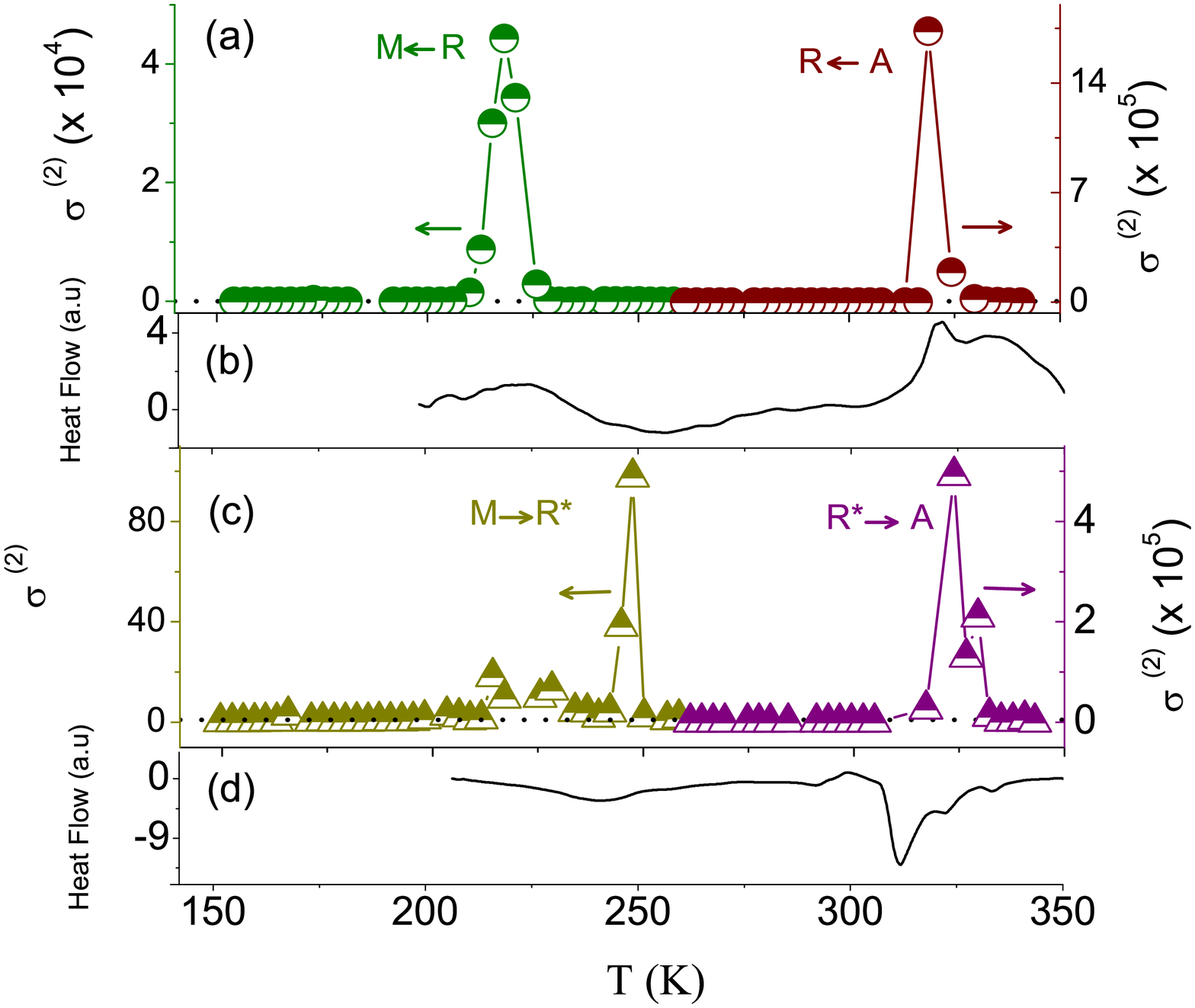}
\end{center}
\vspace{-0.7cm} \caption{(Color Online) The integrated value of the normalized second spectra $\sigma^{(2)}$ as a function of temperature
showing the peaks corresponding to the transitions for (a) cooling and (c) heating. The corresponding DSC plots are shown in the lower panels
(b) and (d). The calculated Gaussian value integrated over the frequency bandpass is shown as a dotted line. The details about the transforming
phases are denoted in (a) and (c).}

\label{figure4}
\end{figure}

From the perspective of material characterization, resistance measurements have long been used as a convenient probe to measure the phase
transformation temperature scales. In case of martensites, these include $M_s$ (martensite start), $M_f$ (martensite finish), $A_s$ (austenite
start), $A_f$ (austenite finish) etc. However, as can be seen from Fig.~5, where $\rho - T$ data from three NiTi wires are compared, it often
becomes very difficult to assign an unambiguous start and finish temperature to the phase transformation region. Instead, by comparing the range
of $T$ in DSC data over which the avalanches are observed, to the width of the $\sigma^{(2)}$ peak, an alternative scheme can be suggested. As
can be seen in the context of martensite transition (M$\leftarrow$R) in Fig.~5, the higher and lower $T$ (denoted by the vertical dashed lines)
at which the value of $\sigma^{(2)}$ starts to deviate from the Gaussian expectation could be assigned as $M_s$ and $M_f$ respectively. In
Fig.~5a the $A_s$ and $A_f$ are identified by the vertical dashed lines, locating the R$\leftarrow$A transition. The same scheme can be adopted
for the other transitions as well.

\subsection{Thermal Cycling}

The system was swept through a temperature range of 360~K-120~K. The corresponding resistance vs. temperature plots are as shown in Fig.1(b). It
is clear that on thermal cycling the defect levels increase and stabilize after around 50 thermal cycles. Subsequently, we measured resistance
fluctuations while the system was being driven at 0.15~K/min across the temperature range of 240~K-180~K (cooling). This region is where the
transformation from the R-phase to martensite phase takes place. It was observed that the spectra exhibits $1/f^\alpha$ behavior. The frequency
exponent $\alpha$ for the entire temperature zone for the various cycles is plotted in Fig.6(a), which is found to cluster around 1.2 and does
not show any prominent variation with cycling. Fig.6(a) also shows a clear decrease in the noise magnitude with decrease in temperature. The
noise magnitude however shows no systematic cycle dependence, although some earlier studies indicated a weak increase in its magnitude with
cycling.

\begin{figure}
\begin{center}
\includegraphics[width=8cm,height=6cm]{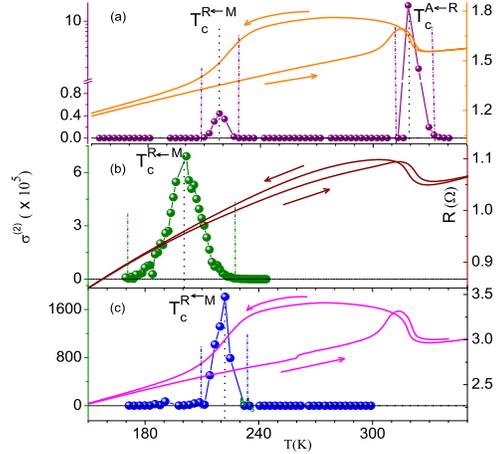}
\end{center}
\vspace{-0.7cm} \caption{(Color Online) Integrated value of the normalized second spectra $\sigma^{(2)}$ as a function of temperature for three
different samples shown along with the corresponding resistance plots. The heating and cooling directions are shown using appropriate arrows.
The region over which $\sigma^{(2)}$ deviates from the Gaussian background (dotted line) is marked using vertical lines. The point at which
$\sigma^{(2)}$ reaches its maximum is defined as the critical temperature $T_c$.} \label{figure2}
\end{figure}

Strikingly however, we observe the critical point $T_c(=T_c^{M\leftarrow R})$ in $\sigma^{(2)}$ shows a clear shift towards lower temperatures
on thermally cycling the sample. To make this behavior clearer, we have plotted $T_c$ as a function of thermal cycles in Fig.~7. To compare with
the changes in the resistance with respect to cycling, we have also plotted the resistance value at 245~K during cooling for as a function of
cycles in the same plot. It is clear that both $T_c$ and resistivity stabilize after similar number of cycles, indicating a common underlying
physics.

\begin{figure}
\begin{center}
\includegraphics[width=9cm,height=8cm]{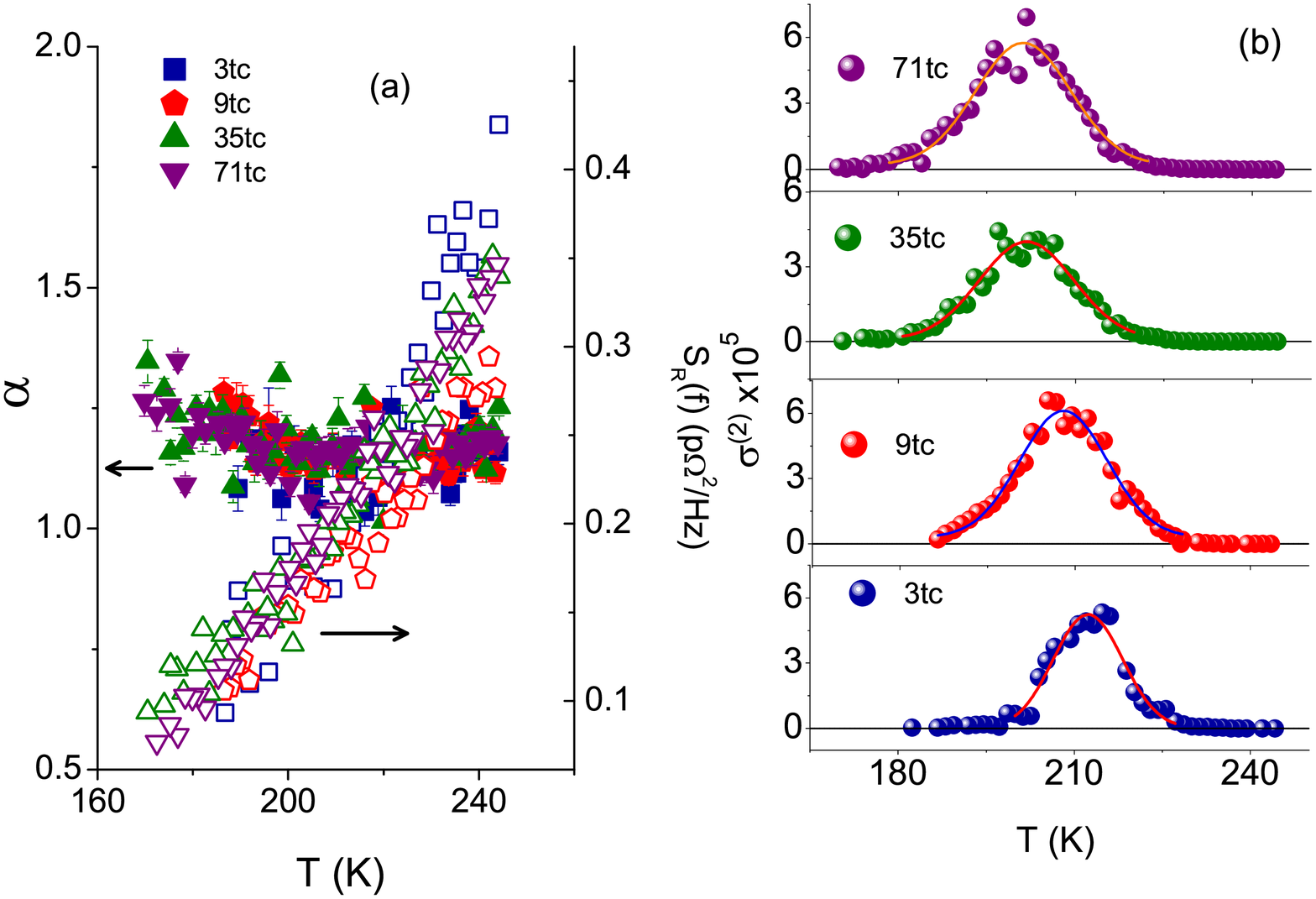}
\end{center}
\vspace{-0.7cm} \caption{(Color Online)(a) Power spectral density (open symbols) and the frequency exponent $\alpha$ (closed symbols) as a
function of temperature for various thermal cycles. (b) The variation of $\sigma^{(2)}$ as a function of temperature for different thermal
cycles. The Gaussian expectation is plotted as dotted lines. The deviation from the Gaussian behavior of the second spectrum around the critical
temperature is evident. The fits shown are Gaussian.} \label{figure3}
\end{figure}

\subsection{Drive rate dependence}
To explore the drive rates effects, the system was driven at five dive rates r
= 0.15~K/min, 0.3~K/min, 0.5~K/min, 0.7K~/min and 1~K/min. To decouple the
effects of drive rates from thermal cycling, this set of experiments was done
only after the disorder levels were stabilized to the equilibrium limit. It was
observed that the peak position $T_c$ for 0.15K/min after and before the drive
rate dependence tests were exactly the same, indicating that the drive rate
experiments have not affected the system's morphology in an irreversible
manner. The PSD does not show rate dependence (Fig.~8(a)), which is essentially
a feature of the athermal nature of the phase transition. The frequency
exponent $\alpha$ is plotted in Fig.8(a), which shows no rate dependence
either.

Fig.~8(c) shows the variation of $\sigma^{(2)}$ for the various rates. As the drive rate increases, there is a gradual shift of $T_c$ towards
lower temperatures. As the drive rate becomes larger, the number of sampling windows decreases, which lead to a larger error bar. In addition,
the magnitude of $\sigma^{(2)}$ is found to increase on increasing the rate. These two features are shown in Fig.~9.

\begin{figure}
\begin{center}
\includegraphics[width=8.5cm,height=7cm]{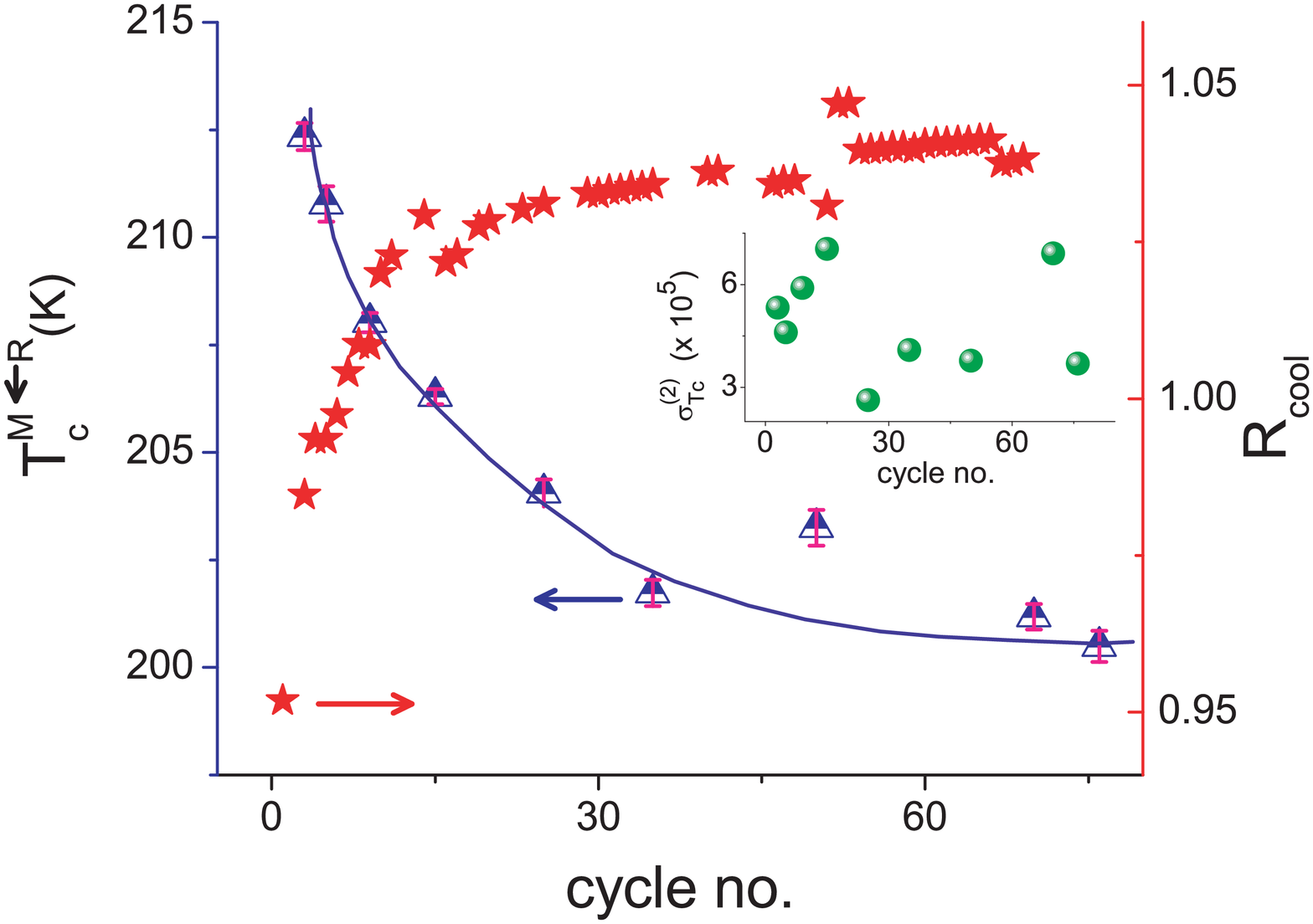}
\end{center}
\vspace{-0.7cm} \caption{(Color Online) The variation of $T_c$ as a function of thermal cycles calculated from the Gaussian fits shown in
Fig.6(b). The curve is a guide to the eye. The resistance value at 245K during cooling for the various thermal cycles is also plotted for
comparison. The inset shows the variation of $\sigma^{(2)}$ at $T_c$ for the various cycles.} \label{figure4}
\end{figure}

\begin{figure}
\begin{center}
\includegraphics[width=9cm,height=7cm]{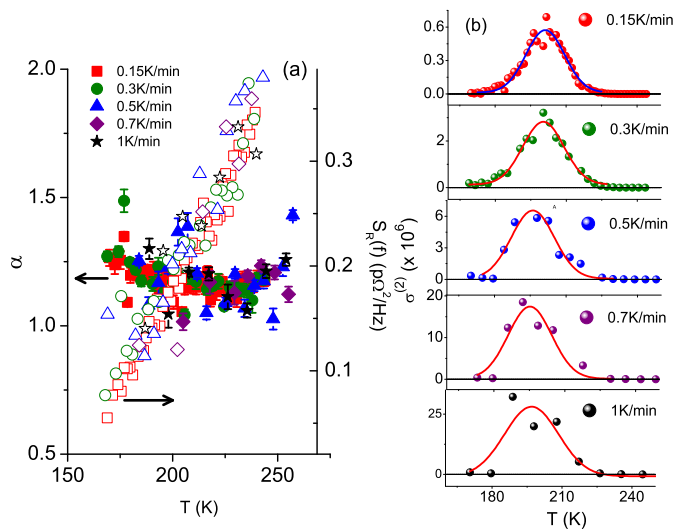}
\end{center}
\vspace{-0.7cm} \caption{(Color Online)(a) Power spectral density (open symbols) and the frequency exponent $\alpha$ (closed symbols) as a
function of temperature for various drive rates. (b) The variation of $\sigma^{(2)}$ as a function of temperature for the various drive rates.
The Gaussian expectation is plotted as dotted lines. The deviation from the Gaussian behavior of the second spectrum around the critical
temperature is evident. The fits shown are Gaussian.} \label{figure5}
\end{figure}

\section{Discussion}
The experimental data indicates that both thermal cycling and drive rates have very similar effects on the behavior of the resistance
fluctuations and its higher order spectra. In this section, we would work to develop a qualitative understanding of these effects based on the
evolution of elastic energy and overlap of avalanches as the system traverses through thermal cycles and the variations in drive rates.

\emph{Thermal Cycling}

The effects of thermal cycling have been extensively investigated by Miyazaki et al. in NiTi samples \cite{miyazaki}. It has been established
that on cycling dislocations are introduced in samples which are not extensively heat treated prior to cycling. This generic behavior has been
characterized using resistance measurements (Fig.1(b)), which shows an increase in resistivity of the wires on thermal cycling due to the
increased number of dislocations. However, the initial increase in dislocations causes an internal strain field which then opposes further
proliferation of dislocations leading to a saturation of disorder level. Perez-Reche et al. have attributed this feature to a self-organization
of defects \cite{reche2}.

\begin{figure}
\begin{center}
\includegraphics[width=9cm,height=7cm]{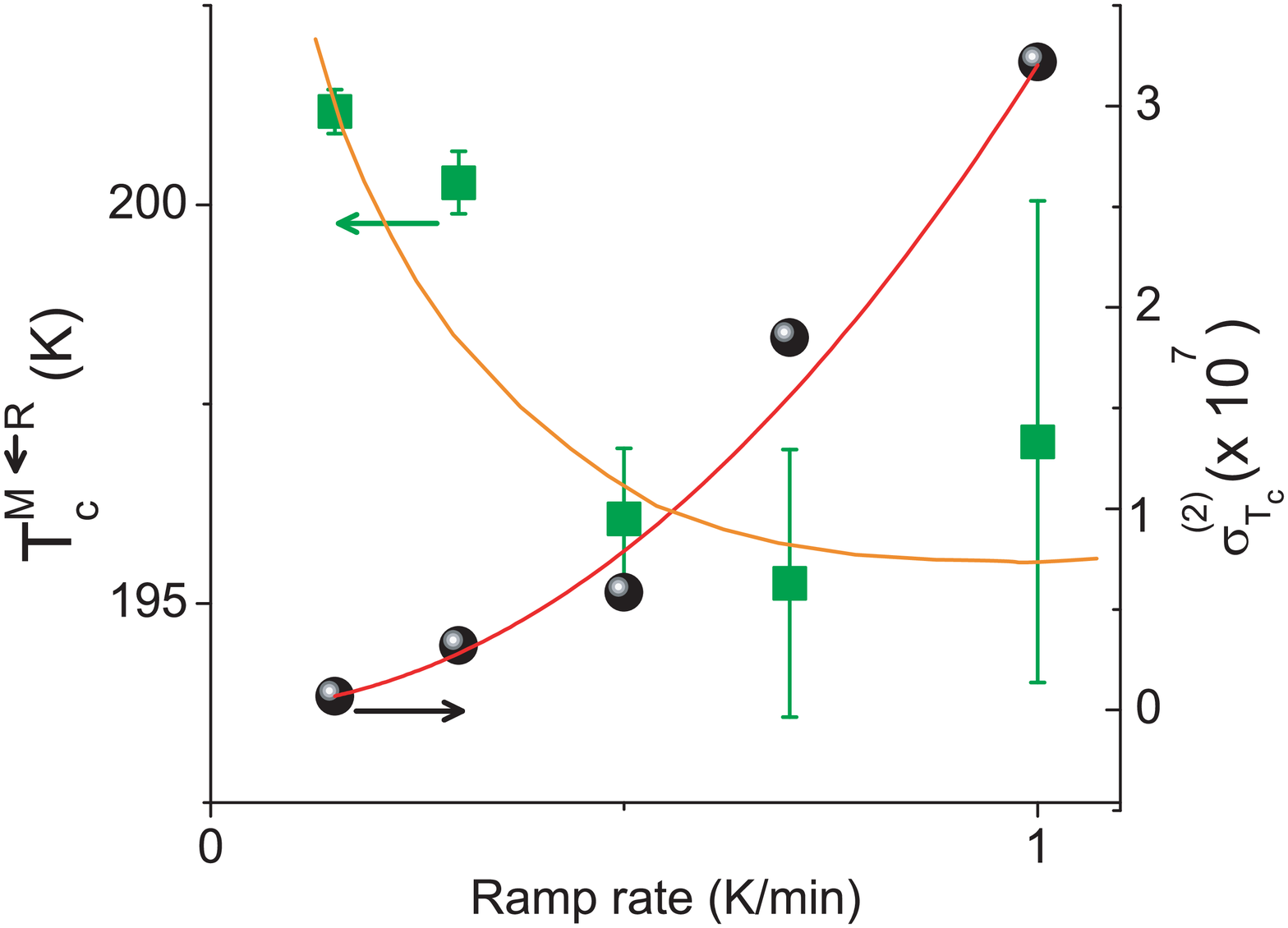}
\end{center}
\vspace{-0.7cm} \caption{(Color Online) The critical point $T_c$ calculated from the Gaussian fits in Fig.8(b), as a function of drive rate. The
magnitude of $\sigma^{(2)}$ at $T_c$ as a function of drive rate is plotted on the right hand side axis. The curves are guide to the eye.}
\label{figure6}
\end{figure}

To explain the shift in $T_c$, it is essential to understand the changes in the internal elastic strain energy and its implication on the
transition temperature scales. From the DSC measurements as well, it has been observed that the martensite start temperature $M_s$ decreases on
thermal cycling. This has been attributed to the increase in dislocation stress field\cite{miyazaki,mccormick}, which results in increasing the
potential energy barrier between the metastable states and hence require an additional under-cooling to drive the phase transformation. This
scenario readily explains the variation in $T_c$ as well, but the absolute magnitude of $\sigma^{(2)}$ was found to be insensitive to thermal
cycling (inset of Fig.7). This implies that while the local elastic energy profiles are modified during thermal cycling, the extent of long
range correlation length remain largely unaffected.

\emph{Drive Rate Effects}

The effect of drive rates has been probed in our earlier work as well \cite{prl} and Fig.8(a) shows very similar results. To understand the
second spectral characteristics (\emph{i.e.} the magnitude of $\sigma^{(2)}$ and the deviation in $T_c$), we adopt a formalism similar to the
one followed in Ref.28.

A driven structural phase transition can be described using three competing timescales: (1) The avalanche relaxation $\tau_{av}$, which depends
on the interaction in the system and the disorder level, being typically $\sim 10^{-8} - 10^{-4}$~sec in most systems. (2) The time scale of
thermal fluctuations ($\tau_{th}$), which depends on the elastic energy barriers separating the metastable states, and depends on the mechanisms
for the transition like slipping, twinning etc. In an athermal transition the effects of thermal fluctuations are small, and $\tau_{th}$ is
generally taken to be very large ($\tau_{th} > 10^5 - 10^6$~sec) in conventional shape memory alloys such at nitinol. (3) The drive rate $r$
itself defines a time scale $\tau_{dr} \propto 1/r$, which is well separated from both $\tau_{th}$ and $\tau_{av}$ in most cases, which
quantifies the athermal limit as $\tau_{th}/\tau_{dr} \rightarrow \infty$, and the adiabatic limit as $\tau_{av}/\tau_{dr} \rightarrow 0$, where
the finite drive does not affect the characteristics and propagation of avalanches.

In a real system both $\tau_{th}$ and $\tau_{av}$ are finite, and at finite drive two processes can occur: (i) the activation of the transition
by thermal fluctuations, and (ii) the overlap of avalanches. The behavior of both $\sigma^{(2)}$ and $T_c$ shown in Fig.~9 can be explained with
these two processes, assuming in general $\tau_{th} > \tau_{dr} > \tau_{av}$ and a scenario depicted schematically in Fig.~10. The vertical
dashed line represents a particular sample with material specific $\tau_{th}$ and $\tau_{av}$. As drive rate increases, $\tau_{av}/\tau_{dr}$
also increases, resulting in larger probability of avalanche overlap. Due to the long range nature of elastic force, superposition of avalanches
would couple events separated in time and space, leading to longer correlations, and hence increase in the magnitude of $\sigma^{(2)}$ as seen
in Fig.~9.

The variation in $T_c$, on the other hand, can be associated to an increasing $\tau_{th}/\tau_{dr}$ when $r$ increases, {\it i.e.} as the system
is {\it effectively} driven towards the athermal limit. Hence the number of smaller avalanches triggered by thermal fluctuations decreases,
requiring an additional undercooling to transform the same amount of the parent phase to product. This naturally leads to a lower value of
$T_c$. It is to be noted that the effect of finite drive rate on $T_c$ would be insignificant if the system is truly athermal
$\tau_{th}/\tau_{dr} \rightarrow \infty$, which is traditionally assumed for the martensite transformation in NiTi~\cite{takeda}. However, our
experiments here indicate that the thermal fluctuations do play a minor role in NiTi. It is also important to note that the power spectral
density of noise changes very little for various drive rates~\cite{prl,acta}, indicating the second spectral properties to be far more sensitive
to small deviations from athermal behavior and adiabatic limit than the noise itself.

\begin{figure}
\begin{center}
\includegraphics[width=8cm,height=7cm]{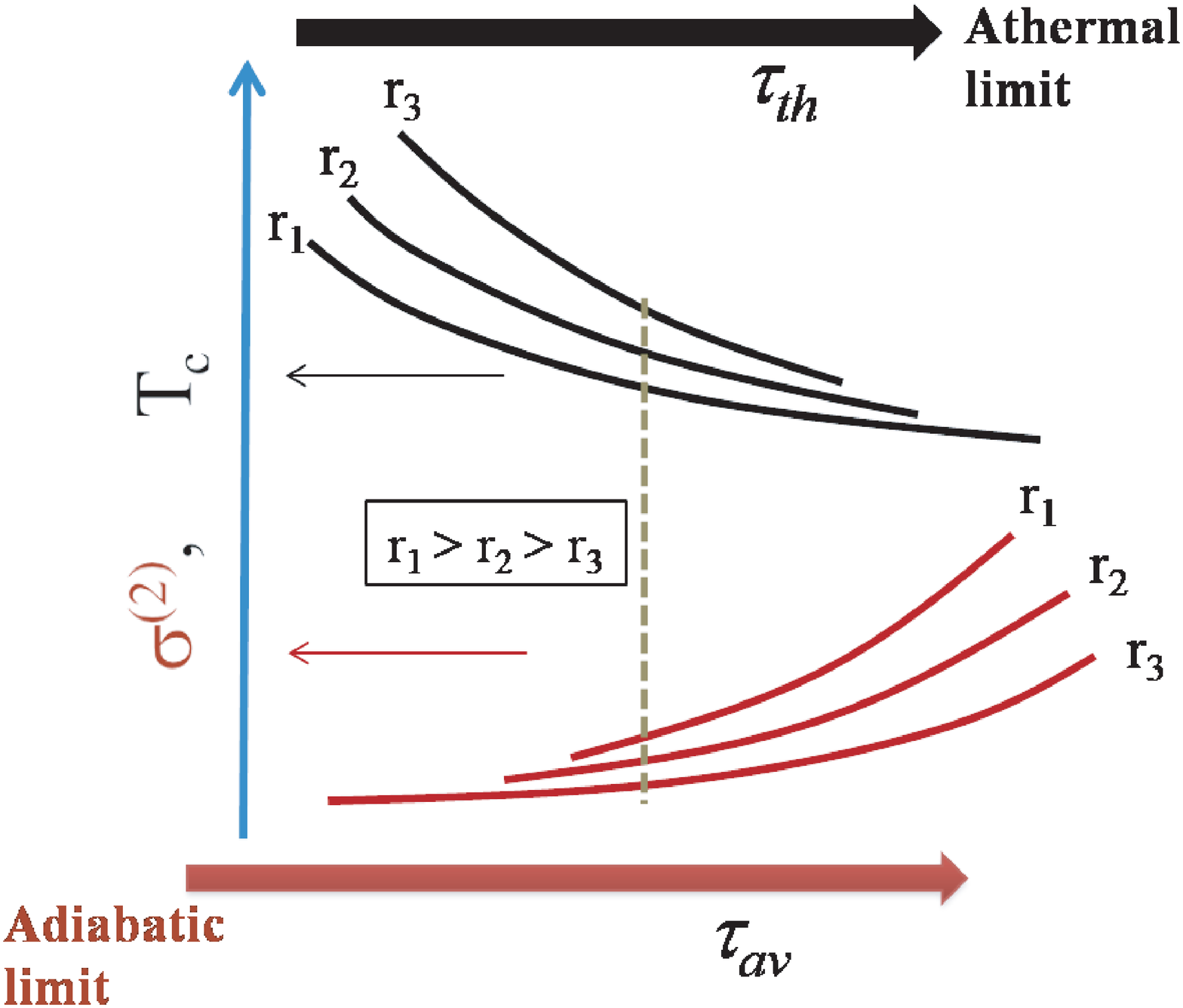}
\end{center}
\vspace{-0.7cm} \caption{(Color Online) Schematic representation of the simplistic model based on the competing timescales, described in the
discussion section. $r_1,r_2$ and $r_3$ denote three different rates.}
\end{figure}

\section{Conclusion}
In this work we have adopted a novel method based on higher order statistics of fluctuations to probe into the kinetics of martensite structural
transformation in nickel-titanium alloys. The non-Gaussian component of noise in electrical resistivity of these systems is found to be an
excellent probe, which relies on the built-in long range elastic correlations in this system. The sensitivity of the technique to small scale
microstructural modifications was also addressed by measuring noise as a function of thermal cycling and the drive rate. We believe this could
serve as an independent and sensitive tool to probe various forms of structural transition in a wide variety of (electrically conducting)
material systems.

\textbf{Acknowledgement} \linebreak U. C. would like to thank Council of Scientific and Industrial Research for a fellowship. A.G thanks Space
Technology Cell, Indian Space Research Organization, for financial support.



\begin{thebibliography}{1}
\bibitem{miguel} M. C. Miguel et al., Nature. \textbf{410}, 667 (2001).
\bibitem{dimiduk} D. M. Dimiduk et al., Science. \textbf{312}, 1188 (2006).
\bibitem{csikor} F. F. Csikor et al., Science. \textbf{318}, 251 (2007).
\bibitem{bradby} J. E. Bradby, J. S. Williams and M. V. Swain, Phys. Rev. B. \textbf{67} 085202 (2003).
\bibitem{otsuka}\textit{Shape Memory Materials}, edited by K. Otsuka and C. M. Wayman, (Cambridge University Press,(1998).
\bibitem{devincre} B. Devicre, T. Hoc and L. Kubin, Science. \textbf{320}, 1745 (2008).
\bibitem{mo} Y. Mo, K. T. Turner and I. Szlufarska, Nature. \textbf{457}, 1116 (2009).
\bibitem{carrillo} L. Carrillo, L. Manosa, J. Ortin, A. Planes and E. Vives, Phys. Rev. Lett. \textbf{81}, 1889 (1998).\bibitem{kuntz} M. C. Kuntz and J. P. Sethna, Phys. Rev. B. \textbf{62}, 11699 (2000).

\bibitem{durin} G. Durin and S. Zapperi, Phys. Rev. Lett. \textbf{84}, 4705 (2000).
\bibitem{sethna1} J. P. Sethna, K. A. Dahmen, S. Kartha, J. A. Krumhansl, B. W. Roberts and J. D. Shore, Phys. Rev. Lett. \textbf{70}, 3347(1993).
\bibitem{sethna2} J. P. Sethna, K. A. Dahmen and C. R. Myers, Nature \textbf{410}, 242 (2001).

2004).
\bibitem{weiss} J. Weiss and J. R. Grasso, J. Phys. B. \textbf{32}, 101 (1997).
\bibitem{reche2} F. J. Perez-Reche, L. Truskinovsky and G. Zanzotto, Phys. Rev. Lett. \textbf{99}, 075501 (2007).
\bibitem{reche} F. J. Perez-Reche, E. Vives, L. Manosa and A. Planes, Phys. Rev. Lett. \textbf{87}, 195701 (2001).
\bibitem{planes} A. Planes, F. J. Perez-Reche, E. Vives and L. Manosa, Scripta Materialia. \textbf{50}, 181 (2004).
\bibitem{juan} J. S. Juan, M. L. No and C. A. Schuh, Nat. nanotech. \textbf{4}, 415 (2009).
\bibitem{prl} U. Chandni, A. Ghosh, H. S. Vijaya and S. Mohan. Phys. Rev. Lett. \textbf{102}, 025701 (2009).
\bibitem{acta} U. Chandni et. al. Acta Materialia. \textbf{57}, 6113 (2009).
\bibitem{vives} E. Vives et. al. Phys. Rev. Lett. \textbf{72}, 1694 (1994).
\bibitem{scofield}\ J. H. Scofield, Rev. Sci. Instrum. \textbf{58}, 985 (1987).
\bibitem{seidler} G. T. Seidler and S. A. Solin, Phys. Rev. B \textbf{53}, 9753 (1996).
\bibitem{restle} P. J. Restle, R. J. Hamilton, M. B. Weissman and M. S. Love, Phys. Rev. B. \textbf{31}, 2254 (1985).
\bibitem{swastik} S. Kar et al. Phys. Rev. Lett. \textbf{91}, 216603 (2003).
\bibitem{hardner} H. T. Hardner et al. J. Appl. Phys. \textbf{81}, 272 (1997).
\bibitem{miyazaki} S. Miyazaki, Y. Igo and K. Otsuka. Acta metall. \textbf{34}, 2045 (1986).
\bibitem{mccormick} P. G. McCormick and Y. Liu. Acta metall. mater. \textbf{42}, 2407 (1994).
\bibitem{reche3} F. J. Perez-Reche et al. Phys. Rev. Lett. \textbf{93}, 195701 (2004).
\bibitem{takeda} K. Otsuka, X. Ren and T. Takeda. Scripta Materialia. \textbf{45}, 145 (2001).
\end{thebibliography}
\end{document}